\documentclass[sigconf,authorversion,nonacm]{acmart}

\usepackage{xcolor}
\usepackage{soul}

\colorlet{usercolorname}{yellow!0}
\sethlcolor{usercolorname}

\usepackage{graphicx}
\usepackage{subcaption}

\usepackage{multirow}
\AtBeginDocument{%
  }

\copyrightyear{2026}
\acmYear{2026}
\setcopyright{cc}
\setcctype{by-nc-nd}
\acmConference[CUI '26]{ACM Conversational User Interfaces 2026}{July 21--24, 2026}{Bremen, Germany}
\acmBooktitle{ACM Conversational User Interfaces 2026 (CUI '26), July 21--24, 2026, Bremen, Germany}
\acmDOI{10.1145/3816046.3816284}
\acmISBN{979-8-4007-2741-2/2026/07}

\begin{document}

\title{Validating the Single Item Kawaii Measure}

\author{Katie Seaborn}
\orcid{0000-0002-7812-9096}
\affiliation{%
  \institution{\small University of Cambridge}
  \city{Cambridge}
  \country{UK}
}
\affiliation{%
\institution{\small Institute of Science Tokyo}
\city{Tokyo}
\country{Japan}
}
\email{katie.seaborn@cst.cam.ac.uk}

\author{Yijia Wang}
\email{wang.y.cf@m.titech.ac.jp}
\orcid{0009-0004-2250-9163}
\affiliation{%
  \institution{Tokyo Institute of Technology}
  \city{Tokyo}
  \country{Japan}
}

\renewcommand{\shortauthors}{Seaborn and Wang}

\begin{abstract}
Kawaii is the Japanese instantiation of cuteness. As a multimodal percept theoretically derived from the notion of baby schema, kawaii can be %but applies to any non-adult and non-human stimuli with infant-like features. Kawaii is a multimodal element, 
a property of voice and sound, visual appearance and form factor, and movement and expression.
% of many popular voice actors, video game characters, and voice assistants. 
However, measuring user perceptions of kawaii remains an open question. In the absence of a validated instrument, a one-item self-report measure has been used extensively, but has not been validated. Here, we report on three types of validity---convergent, known groups, and cross-context---and reliability for the single item measure across \hl{nine} data sets featuring responses to \hl{video game character voices and visual appearances and computer-generated voice assistant voices} from $N=967$ unique participants. Our results demonstrate initial evidence of the validity of the one-item measure for \hl{voice and visual} kawaii perceptions. Further rigour can be pursued with novel stimuli, test-retest validation, and concurrent validity against the upcoming multi-item measure of kawaii.
  % perceptions of cuteness and a caregiving response to infantile stimuli. 
  
\end{abstract}





\maketitle

%% ------------------------------------

\section{Introduction}
\label{sec:intro}

Kawaii translates to ``cute'' in Japanese, but carries sociocultural meanings and instantiations that nevertheless transcend the borders of Japan. Simply put, kawaii is a form of baby schema translated into non-baby forms that elicits certain socioaffective reactions in people and other animals~\cite{nittono_power_2012,nittono_psychophysiological_2017,nittono_two-layer_2016}. Japan is well-known for its kawaii-ified exports, from video game characters to pop idols to anime voice actors. As a sociocultural phenomenon, kawaii is normalized in Japan, even while kawaii is leveraged as design material in various media. Kawaii voices, in particular, have emerged as a (re)new(ed) modality in the human-computer interaction (HCI) domains of human-agent interaction (HAI) and human-AI interaction (HAII) through work on voice assistants (VAs), conversational user interfaces (CUIs), and virtual characters of all kinds. Notably, Nittono et al.~\cite{nittono_power_2012,nittono_two-layer_2016} developed a general science of kawaii, later extended by Seaborn et al.~\cite{seaborn_can_2023,seaborn_kawaii_2023,mandai_super_2025} to encompass kawaii vocalics (the vocal medium and paralanguage), with both groups of researchers applying the concept to a range of computer-based agents, including social robots, VAs, and video game characters. Kawaii has been linked to a range of user experience (UX) factors---like  emotion~\cite{nittono_psychophysiological_2017}, trust~\cite{Shiomi_2023}, performance~\cite{yoshikawa_effects_2020}, persuasion~\cite{nittono_power_2012,birlea2023soft}, deception~\cite{zhang_first_2025}---positioning it as an important design(able) element that deserves careful consideration~\cite{wang2024kawaiicomp,Ohkura_2023kawaiiengineering}.

Yet, how to evaluate the UX of kawaii---represented in visual or vocal stimuli---remains an open question. Given its socioaffective baseline, subjective self-reports, observations, and physiological measures are most viable~\cite{nittono_two-layer_2016}. However, these modes of evaluation may not be viable in all cases. The move to online studies following the COVID-19 global pandemic~\cite{Costagliola2023} makes physiological measurement difficult. Contexts like in situ studies for longitudinal evaluation limit the forms of observation possible~\cite{vanMechelen1997}. Self-report formats seem ideal, as these can be integrated into a range of online, asynchronous engagement formats, notably questionnaires.

Even so, few self-report methods for kawaii exist~\cite{wang2024kawaiicomp}. Those that do bear limitations for HCI work. 
% Importantly, the validated, multi-item instruments may not translate to computer media. 
For example, \citet{Takamatsu2018} developed the 15-item Cuteness Responsiveness (CR-15) self-report scale, focusing on caregiving responses to human and non-human animals and mascots. However, the scale measures general trait-level responses. Moreover, the caregiving and baby-centric focus---while common in cute studies~\cite{Glocker2009}---does not necessarily transfer. For instance, one positive item reads ``If a baby cries, I want to approach and soothe her'' (a caregiving response) and one negative item is ``It’s terrible to see stray cats and dogs'' (bearing an unknown relationship to cuteness, since strays can still be cute, and geared more towards eliciting or not eliciting a caregiving response). 
%Finally, the scale was only validated with a typical Japanese parent-child cohort. 
% https://link.springer.com/article/10.1007/s12144-018-9836-4/tables/1
This led to the emergence of a single-item measure for user perceptions of kawaii~\cite{wang2024kawaiicomp}, where participants are asked to rate ``kawaii'' as an attribute of a stimulus in a Likert-scale response. This measure, normally folded into other instruments or alongside other descriptive items, like ``enjoyable'' and ``trustworthy,'' has been applied to computer agents ranging from VAs~\cite{mandai_super_2025,seaborn_can_2023} to video game characters~\cite{seaborn_kawaii_2023}.
Yet, the measure is not validated. Previous research, notably in the CUI space and specifically relevant to vocalics, has raised the problem of over-reliance on %novel measures and/or 
unvalidated single-item measures~\cite{seaborn_voice_2025, seaborn_voice_2021,seaborn_measuring_2021}. Essentially, this is a matter of validity~\cite{Allen2022single}: the item may not measure what we think it is measuring, or may not be sufficient on its own to measure the phenomenon under study, or may not transfer across people, contexts, and instantiations.
% : visual vs. vocal, CUIs vs. NPCs (non-playable characters).

In response, we gathered together the data sets including the single-item measure of kawaii and conducted a validity analysis. Our goal was to ascertain the degree of validity and reliability of the measure, with the secondary goal of assessing the rigour of the research so far. We asked: \emph{\textbf{Is the single-item measure of kawaii valid?}} We provide initial evidence of the item's validity across different validity types, kawaii stimuli modality (visual and voice), and platform (VAs, video game characters, and recorded voices). We contribute baseline confidence in prior and continued use of the single-item measure of kawaii, an important finding while a multi-item scale remains absent.

%% ------------------------------------

\section{Theoretical Background}
\label{sec:bg}

% Kawaii is the Japanese instantiation of cuteness. 
The baseline theory is baby schema or Kindchenschema~\cite{Lorenz1943}. Baby schema is a model of infant features: a large head in relation to the body; large eyes, big cheeks; small nose; high forehead; general roundness. These features of the smallest and most vulnerable members of the species invoke certain socioaffective reactions. Primary is a caregiving response~\cite{nittono_two-layer_2016,nittono_psychophysiological_2017,Glocker2009,Takamatsu2018}. But kawaii also elicits feelings of being charmed and disarmed, smiles, mannerisms, and even social bonding, like the ``kawaii triangle'' that emerges among those gathered around a cute stimulus~\cite{nittono_two-layer_2016}.

The Japanese sociocultural context adds a further layer: cuteness is normalized in company logos, fashion, and exports like anime and video game characters loved by people of all ages. Hence, when studying cuteness within Japanese contexts (e.g., reactions to vocaloids) or with people raised in Japanese culture, understanding and considering the kawaii perspective is critical. Notable are sociocultural behavioural responses like kawaii spirals and triangles, culturally relevant phenomena like chizimi shikou (love of small things), and amae, or the desire to gain other people's affections~\cite{nittono_power_2012,nittono_two-layer_2016}. Still, these apparently Japan-specific factors may transfer to other cultures; more research is needed.

Most theory and work has focused on visual modalities and appearance, e.g., \citet{nittono_two-layer_2016,Glocker2009,Takamatsu2018,Ohkura_2023kawaiiengineering}. \citet{nittono_two-layer_2016} developed a two-layer model of kawaii that encompasses kawaii as an emotion and a social value, i.e., socioaffective. The model links attributes like baby schema, smiling, roundness, and colour to a stimulus, which is perceived as ``cute,'' ``friendly,'' ``harmless,'' and so on (cognitive appraisals), leading to the positive emotion of kawaii, which manifests in subjective, behavioural, and/or physiological responses. 
Likewise, \citet{Ohkura_2023kawaiiengineering} and contributors identified several variables relevant to artificial kawaii products (computers but also characters), including colour, shape, material, sound, and various emotional responses. Importantly, kawaii/cuteness is also associated with sound and notably voice~\cite{wang2024kawaiicomp,seaborn_kawaii_2023,seaborn_can_2023,mandai_super_2025}. \citet{seaborn_can_2023} proposed that kawaii voice stimuli elicit similar reactions to visual stimuli. Moreover, they leveraged previously unacknowledged (and seemingly normalized) linkages between social characteristics and kawaii, notably gender~\cite{Shiomi_2023,BURDELSKI2010,AsanoCavanagh2014,Iseri2015} and age~\cite{Ota2022,lieber2021cute,Kroo2018}, confirming the relevance to kawaii: young age and feminine and ambiguous genders. They further contextualized this kawaii vocalics model for computer voice stimuli, identifying humanlikeness and fluency as key factors. This first effort, applied to VAs, was replicated for video game characters~\cite{seaborn_kawaii_2023,mandai_super_2025}. From this work and the associated data sets, we identified six cross-modal factors as dimensions of kawaii---\emph{humanlikeness}~\cite{seaborn_can_2023,seaborn_kawaii_2023,mandai_super_2025}, \emph{lack of artificiality/machinelikeness}~\cite{seaborn_can_2023,seaborn_kawaii_2023,mandai_super_2025}, \emph{happiness}~\cite{nittono_two-layer_2016,Ohkura_2023kawaiiengineering,mandai_super_2025}, \emph{trustworthiness}~\cite{mandai_super_2025}, \emph{favourableness}~\cite{Ohkura_2023kawaiiengineering,mandai_super_2025}, and \emph{excitement}~\cite{Ohkura_2023kawaiiengineering,mandai_super_2025}.

%% Yijia, I feel you copy-pasted this from somewhere. I asked you to simply add a list of factors. There's not enough room and it doesn't fit. So I had to re-write it :\
% Prior work provided theoretical and statistical evidence of factors related to kawaii perception. \citet{seaborn_can_2023} first found statistically positive association between humanlike-ness and voice kawaii-ness, and negative association between artificiality and voice kawaii-ness. Subsequent studies have shown that these findings extend to a broad range of stimuli, such as game character voices~\cite{seaborn_kawaii_2023, mandai_super_2025}. \citet{nittono_two-layer_2016} suggested that kawaii is a positive emotion, while \citet{Ohkura_2023kawaiiengineering} identified emotion-related variables (e.g. happiness and pleasantness) associated with visual kawaii.\citet{mandai2025super} extended this association to the auditory domain, and found that perceived voice kawaii-ness is positively associated with happiness, trustworthiness, favorability, and excitement.
% Guided by these findings, we proposed the theoretically derived initial multi-item scale---\emph{Humanlike, Artificial/Machinelike, Happy, Trustworthy, Favourable, and Excited}---served as external validation criteria for evaluating the validity of the single-item kawaii measure.

Kawaii is a subjective, socioaffective response to a stimulus with features linked to baby schema and cultural elements: in short, a (user) percept. Most work has focused on caregiving responses~\cite{yoshikawa_effects_2020,Glocker2009,nittono_two-layer_2016}, with only some on more ``artificial'' applications of cuteness and kawaii~\cite{nittono_power_2012,wang2024kawaiicomp,Ohkura_2023kawaiiengineering} despite a long history as design material, as discussed. Now, VAs, CUIs, and interactive characters/virtual humans are being designed with kawaii in voice, body, and mannerisms. %Perhaps due to positivity biases, m
Work has targeted negative applications, like conniving home robots that reduce user agency over time~\cite{Lacey2019} and video game characters that lure players into spending more attention or money~\cite{zhang_first_2025}. The evolutionary caregiving-oriented forms of measurements are also not applicable. This led to the emergence of the one-item measure for computer-based stimuli~\cite{mandai_super_2025,seaborn_can_2023,seaborn_kawaii_2023}. When no established model or validated instrument exists, single item measures can be appropriate---or the only option~\cite{Allen2022single}. They are also parsimonious: answering one item, especially amidst other scales, is less burdensome for the respondent~\cite{Allen2022single} and data analyst~\cite{Allen2022single,Bergkvist2007}, and might even attract respondents to future studies~\cite{Wanous1997}. When the factor is well-defined and circumscribed, like kawaii, a one-item measure may be ideal~\cite{Allen2022single, fuchs2009using}. Still, the complexity of kawaii for a single stimulus---given the noted theoretical advances---deserves consideration. A multi-item instrument is under development. Here, we aimed to discover whether the single item measure captures the complexity of kawaii as a percept in a valid and replicable way.
% , until and after the publication of the multi-item instrument.

% \hl{Theories of Kawaii and Baby Schema ... just the two models, quickly}

%% ------------------------------------

\section{Methods}
\label{sec:methods}

We assessed the validity of the single item measure using existing data, following \citet{Allen2022single}. 
Given the nature of the data sets available, we could not use all validation means. There is no minimum for validity, but ideally all forms should be pursued, if possible~\cite{Allen2022single,Boateng2018}.
\citet{Allen2022single} suggests criterion validity through convergent validity, or comparison to the multi-item version of the measure or theoretically-linked items.
\citet{Frongillo2019} further proposes cross-context validity, where the same measure is applied to different stimuli, modalities, or environments. \citet{Boateng2018} also offer construct validity via known groups that differ in expected ways.
In this work-in-progress, we used convergent validity (comparing the single item to multiple items), differentiation by known groups in kawaii vs. non-kawaii stimuli (as construct validity), and cross-context validity by modality (voice vs. body) and cohort (voice only).

\subsection{Data Sets}
We used open data or requested data sets that used the same one-item measure of kawaii (\autoref{tab:data}). However, we did not use the ``super kawaii''~\cite{mandai_super_2025} manipulated data to avoid noise due to perceptual contamination. The data comprised $N=1228$ total and $N=967$ unique participants within Japan recruited through the high-quality and representative platform Yahoo! Crowdsourcing~\cite{seaborn_quality_2025}. The data sets can be accessed here: 
\url{https://bit.ly/validatingkawaii}

\begin{table*}[!ht]
\caption{Data sets used.}
\label{tab:data}
\begin{tabular}{rrlllll}
\toprule
ID & Year & Stimuli & Media & Target & $N$ & Source \\
\midrule
D1 & 2022 & Voice Assistants & Voice & Voice & 94 & \citet{seaborn_can_2023}   \\
D2  & 2023 & Video Game Characters & Voice & Voice & 157 & \citet{seaborn_kawaii_2023}   \\
D3  & 2024 & Video Game Characters & Voice & Voice & 150 & N/A   \\
D4  & 2024 & Video Game Characters & Body & Body & 51 & N/A   \\
D5  & 2024 & Video Game Characters & Voice+Body & Voice & 130 & N/A   \\
% D6  & 2024 & Video Game Characters & Voice+Body & 131 & N/A   \\
D7  & 2024 & Video Game Characters & Voice+Body & Both & 261 & N/A   \\
D8  & 2024 & Video Game Characters & Voice+Body & Voice & 154 & N/A   \\
D9  & 2023 & Voice Assistants & Voice & Voice & 50 & \citet{mandai_super_2025}   \\
D10 & 2024 & Video Game Characters & Voice & Voice & 51 & \citet{mandai_super_2025}  \\
\bottomrule
\end{tabular}
\end{table*}

\subsection{Data Analysis}

% Face validity can be assessed in different ways~\cite{Connell2018}. The goal is to determine how understandable and relevant the item is. In the absence of qualitative means, comparisons across stimuli that are expected to be kawaii or not is appropriate.
Convergent validity was assessed by comparing the single item measure to items considered for the in-progress multi-item instrument (D2--5, D7, D8). Cronbach's $\alpha$ (alpha)~\cite{Cronbach1951} was used for reliability~\cite{Boateng2018}, with $\alpha$ values .70--.95 deemed acceptable~\cite{Tavakol2011}.
Validation was made via Kendall's $\tau_b$ (tau-b), which is appropriate for Likert-scale and non-normal data~\cite{vandenHeuvel2022}.
We interpreted the $\tau_b$ for rank-based correlation statistics via \href{https://blogs.sas.com/content/iml/2023/04/05/interpret-spearman-kendall-corr.html}{Wicklin} by \citet{Schober2018}: $\tau_b=.00$ as negligible, $\tau_b=.06$ as weak, $\tau_b=.26$ as moderate, $\tau_b=.49$ as strong, and $\tau_b=.71$ as very strong.
% Given no guidance for the $\tau_b$ but its relevance to Person's $r$~\cite{vandenHeuvel2022}, we applied the interpretations for $r$ by \citet{Allen2022single}: $r=.90$ as excellent, $r=.80$ as good, $r=.70$ as acceptable, and $r=.60$ or less questionable.
% in which r = .90 is indicative of excellent convergent validity, r = .80 indicates good convergent validity, r = .70 indicates acceptable convergent validity, r = .60 indicates questionable convergent validity, and r < .60 indicates poor convergent validity (Greiff & Allen, 2018).

For differentiation by known groups, we used Kendall's $\tau_b$, comparing kawaii vs. non-kawaii stimuli (D1). We could not use M5 from D9 because the sample size was insufficient.

We used two approaches to cross-context validity with Kendall's $\tau_b$. We compared voice (D10) and body (D4);
% Cross-context validity involved comparing the same source by voice (D2--3, D10) and body (D4) 
% KS: NO ROOM
%across cohorts (D1 v. D9, comparing D2--3, D5, D10)  
we could not use D2 and D3 because of extreme sample size differences.
We also compared voice by cohort (D2 and D3); again, due to sample sizes, we could not compare D1 and D9.

%% ------------------------------------

\section{Results}
\label{sec:results}

\subsection{Convergent Validity}

We compared the single item to multiple items within each data set (D2--5, D7, D8); refer to \autoref{tab:converg}. Reliability via Cronbach's $\alpha$ was acceptable, supporting baseline item interrelatedness and the single item's relevance to the other items. Kendall's $\tau_b$ tests indicated statistically significant relationships between the single item and the items in the instrument, with all correlations strong or very strong. This means excellent convergent validity with a shared theoretical baseline.

\begin{table*}[!ht]
\caption{Convergent validity results. Sig.: Statistically significant. $\alpha$: Cronbach's alpha. $\dag$: Item reversed. *** \emph{p} \textless 0.001.}
\label{tab:converg}
\begin{tabular}{llrlrlrrrrl}
\toprule
\textbf{Data} & \textbf{Target} & \textbf{Scale $\alpha$} & \textbf{Highest}  & \textit{\textbf{$\alpha$}} & \textbf{Lowest} & \textit{\textbf{$\alpha$}} & \textit{\textbf{N}} & \textit{\textbf{$\tau_b$}} & \textit{\textbf{p}} & \textbf{Sig.} \\
\midrule
D2    & Voice   & 0.757   & Happy  & 0.766 & Favourable & 0.688 & 2826   & 0.489 & \textless 0.001     & ***   \\
D3    & Voice   & 0.781   & Artificial$\dag$  & 0.783 & Favourable & 0.703 & 2700   & 0.598 & \textless 0.001     & ***   \\
D4    & Body    & 0.653   & Humanlike & 0.799 & Favourable & 0.526 & 918 & 0.510 & \textless 0.001     & ***   \\
D5    & Voice   & 0.658   & Machinelike$\dag$ & 0.804 & Excited & 0.705 & 650 & 0.559 & \textless 0.001     & ***   \\
D7    & Both    & 0.779   & Machinelike$\dag$ & 0.797 & Favourable & 0.703 & 522 & 0.502 & \textless 0.001     & ***   \\
D8    & Voice   & 0.793   & Artificial$\dag$  & 0.788 & Excited & 0.738 & 770 & 0.486 & \textless 0.001     & ***  \\
\bottomrule
\end{tabular}
\end{table*}

%% ------------------------------------
\subsection{Construct Validity}

We used differentiation by known groups: kawaii vs. non-kawaii stimuli (\autoref{tab:construct}). We used the highest and lowest kawaii and non-kawaii voices from D1.
% Kendall's $\tau_b$ test results are presented in . 
The expected positive correlations among kawaii voices were statistically significant and moderate--strong. While the negative correlations with non-kawaii voices were not statistically significant, this may make sense. While we cannot be fully sure how the response scale was interpreted, it is reasonable to assume that the neutral centre means ``not applicable'' and negative scores mean ``uncute'' (the opposite of kawaii). Hence, the tendency towards the neutral centre is in line with the idea that non-kawaii voices not necessarily ``uncute''---just that cuteness is not applicable.

\begin{table*}[!ht]
\caption{Construct validity results. Sig.: Statistically significant. ** \emph{p} \textless 0.01. *** \emph{p} \textless 0.001.}
\label{tab:construct}
\begin{tabular}{llll|llrrrrrl}
\toprule
\textbf{Kawaii} & \multicolumn{1}{r}{\textbf{M}} & \multicolumn{1}{r}{\textbf{MD}} & \multicolumn{1}{r}{\textit{\textbf{N}}} & \textbf{Comparison} & \textbf{Not} & \textbf{M} & \textbf{MD} & \textit{\textbf{N}} & \textit{\textbf{$\tau_b$}} & \textit{\textbf{p}} & \textbf{Sig.} \\
\midrule
Nana & \multicolumn{1}{r}{3.9} & \multicolumn{1}{r}{4.0}  & \multicolumn{1}{r|}{94} & Not    & M3    & 1.4 & 1.0  & 94 & -0.140 & 0.139  &  \\
& & & & & Okoru & 1.4 & 1.0  & 94 & -0.124 & 0.189  &  \\
& & & & & M2    & 1.5 & 1.0  & 94 & -0.156 & 0.094  &  \\
& & & & Kawaii & Kenshin & 3.9 & 4.0  & 144    & 0.345  & \textless 0.001 & ***    \\
& & & & & Sakura & 3.7 & 4.0  & 94 & 0.461  & \textless 0.001 & ***    \\
\midrule
Kenshin  & \multicolumn{1}{r}{3.9} & \multicolumn{1}{r}{4.0}  & \multicolumn{1}{r|}{144}    & Not    & M3    & 1.4 & 1.0  & 94 & 0.793  & 0.025  &  \\
& & & & & Okoru & 1.4 & 1.0  & 94 & 0.006  & 0.953  &  \\
& & & & & M2    & 1.5 & 1.0  & 94 & 0.024  & 0.795  &  \\
& & & & Kawaii & Sakura & 3.7 & 4.0  & 94 & 0.279  & 0.002  & ** \\
\midrule
Sakura   & \multicolumn{1}{r}{3.7} & \multicolumn{1}{r}{4.0}  & \multicolumn{1}{r|}{94} & & M3    & 1.4 & 1.0  & 94 & -0.074 & 0.436  &  \\
& & & & & Okoru & 1.4 & 1.0  & 94 & -0.131 & 0.166  &  \\
& & & & & M2    & 1.5 & 1.0  & 94 & -0.138 & 0.138  & \\
\bottomrule
\end{tabular}
\end{table*}

%% ------------------------------------
\subsection{Cross-Context Validity}
% We compared by modality across data sets---voice (D2--3, D10) vs. body (D4)---by stimulus.
We compared by modality, using voice data in D10 and body data in D4 from 18 stimuli. A statistically significant, positive but weak correlation was found, $N=918, \tau_b=0.069, p=0.012$. This suggests some disconnect between voice and body versions of the measure, even while a core percept connects the two. Still, characteristics like kawaii may be designed or perceived differently by modality, even for one stimulus (e.g., a robot voice not matching its robot body~\cite{McGinn2019}).
% Follow-up exploratory analyses by individual stimulus revealed could be traced to differences in the source stimuli.
We also compared voice by cohort (D2 at $N=2826$, D3 at $N=2700$). A statistically significant, positive correlation was found, $\tau_b=0.200, p<0.001$, which is considered moderate. This suggests that different people understand the kawaii measure for voice perceptions in a relatively similar way. Overall, this demonstrates good cross-context validity.

%% ------------------------------------
% KS: NO ROOM
% We then compared by cohort across data sets by stimulus: the voice assistants in D1 and D9 and the video game voices in D2--3, D5, and D10.
% \hl{tbd}

%% ------------------------------------

\section{Discussion, Limitations, \& Conclusion}
\label{sec:discuss}

Altogether, the single item measure for kawaii bears good initial validation. This result confirms baseline rigour in the work so far. Moreover, this finding supports continued use of the measure as a parsimonious option for self-reporting perceptions of kawaii---which may be a significant contribution in itself, especially when multiple measures are captured and the burden on participants and analysts is high~\cite{Allen2022single}. The findings also have relevance for future systematic meta-analyses, given that the measure seems valid and has significant use in foundational work.

Still, we were not able to validate the measure in all possible ways, limiting the scope of validation. This can be addressed with further data collection (the sample in D9, for instance, was too small), in line with \citet{Allen2022single} and \citet{Boateng2018}. \citet{Allen2022single} propose further means of validation: face validity, or the pertinence to the respondent~\cite{Mosier1947}; criterion validity in terms of predictive validity---correlating responses to theoretical outcomes---and concurrent validity---high predictive validity compared to the multi-item version; and test-retest reliability, when the measure is expected to be measurable and stable over time. \citet{Boateng2018} offer an overall approach to developing an instrument, starting with expert-driven consensus on baseline items up to validation. Future work should also include ``uncute'' stimuli to resolve the construct validity results. Finally, the data sets involved multiple ratings from the same rater on different stimuli, which could inflate the results. The overall picture is consistent, but future work could evaluate setups of one response to one stimulus per participant.

The next steps are clear. A multi-item instrument applicable to user perceptions of kawaii agents, characters, and voices is underway. The one-item measure should be validated against this instrument. Further validation and assessments of reliability will also be needed, like with non-Japanese samples and materials, as well as a greater range of agent form factors, like social robots. Given the cultural grounding of kawaii alongside its uptake as a factor of study in non-Japanese HCI contexts~\cite{wang2024kawaiicomp}, re-validation of the one-item measure with non-Japanese cohorts will be necessary. Finally, if possible, the measure could be correlated against existing instruments, like the CR-15~\cite{Takamatsu2018}, if an appropriate context of use is found, like parenting reactions to baby robots in distress.

%% ------------------------------------
% \yw{You can use} \verb|\yw| \yw{to make comments}

% \begin{figure}[h]
%   \centering
%   \includegraphics[width=\linewidth]{sample-franklin}
%   \caption{1907 Franklin Model D roadster. Photograph by Harris \&
% Ewing, Inc. [Public domain], via Wikimedia
% Commons. (\url{https://goo.gl/VLCRBB}).}
%   \Description{A woman and a girl in white dresses sit in an open car.}
% \end{figure}

\begin{acks}
Thank you to our colleagues for contributing to kawaii research and notably providing the data sets used. This work was funded in part by a Japan Society for the Promotion of Science (JSPS) Grant-in-Aid for Scientific Research B (KAKENHI Kiban B) grant (no. 24K02972).
\end{acks}

\bibliographystyle{ACM-Reference-Format}
\balance
\bibliography{REFS}

% \appendix









\end{document}